# The Halos of Planetary Nebulae in the Mid-Infrared: Evidence for Interaction with the Interstellar Medium

G. Ramos-Larios & J.P. Phillips

Instituto de Astronomía y Meteorología, Av. Vallarta No. 2602, Col. Arcos Vallarta, C.P. 44130 Guadalajara, Jalisco, México   e-mails : jpp@astro.iam.udg.mx, gerardo@astro.iam.udg.mx

**Abstract**

The motion of planetary nebulae (PNe) through the interstellar medium (ISM) is thought to lead to a variety of observational consequences, including the formation of bright rims; deformation and fragmentation of the shells; and a shift of the central stars away from the geometric centres of the envelopes. These and other characteristics have been noted through imaging in the visual wavelength regime. We report further observations of such shells taken in the mid-infrared (MIR), acquired through programs of IRAC imaging undertaken using the *Spitzer* Space Telescope (SST). NGC 2440 and NGC 6629 are shown to possess likely interacting halos, together with ram-pressure stripped material to one side of their shells. Similarly, the outer halos of NGC 3242 and NGC 6772 appear to have been fragmented through Rayleigh-Taylor (RT) instabilities, leading to a possible flow of ISM material towards the inner portions of their envelopes. If this interpretation is correct, then it would suggest that NGC 3242 is moving towards the NE; a suggestion which is also supported through the presence of a 60 $\mu$m tail extending in the opposite direction, and curved bands of H$\alpha$ emission in the direction of motion – components which may arise through RT instabilities in the magnetized ISM.

NGC 2438 possesses strong scalloping at the outer limits of its AGB halo, probably reflecting RT instabilities at the nebular/ISM interface. We also note that the interior structure of the source has been interpreted in terms of a recombining shell; an hypothesis which may not be consistent with central star luminosities.

Finally, we point out that two of the rims (and likely shock interfaces) appear to have a distinct signature in the MIR, whereby relative levels of 8.0 $\mu$m emission are reduced. This may imply that the grain emission agents are depleted in the post-shock AGB regimes.

**Key Words:** (ISM:) planetary nebulae: individual: NGC 2438, NGC 2440, NGC 3242, NGC 6629, NGC 6772 --- ISM: jets and outflows --- infrared: ISM --- instabilities --- magnetic fields

# 1. Introduction

Deep visual imaging of the shells of planetary nebulae (PNe) has shown that many (perhaps most of them) consist of tripartite structures: an inner bright rim caused by interaction between the stellar wind and AGB envelope; a less bright and but more extended shell, possibly arising from early expansion of D-type ionisation fronts; and very much larger, fainter halos, likely the result of "flash" ionisation of the AGB mass-loss envelope (see e.g. the images and discussion in Corradi et al. 2003, and theoretical analysis by Schönberner & Steffen 2002; Steffen & Schönberner 2003; and Schönberner et al. 2005).

The availability of high quantum efficiency detectors in the 1970s permitted us to determine that whilst many of the halos were circularly symmetric, a good proportion showed evidence for asymmetries, and displacements between the geometrical centres of the envelopes and the positions of the central stars (CSs). Such asymmetries were quickly attributed to interaction of the PNe with the interstellar medium (ISM), and studies of the structures of these sources by Tweedy and others (e.g. Borkowski, Tsvetanov, & Harrington 1993; Tweedy & Napiwotzki 1994; Tweedy & Kwitter 1994a, b; Jacoby & Van de Steene 1995; Tweedy, Martos & Noriega-Crespo 1995; Hollis et al. 1996; Xilouris et al. 1996; Zucker & Soker 1997) culminated in the Atlas of interacting PNe by Tweedy & Kwitter (1996). Several characteristics were cited as evidence for interaction with the ISM, including the afore-stated displacement of the central stars; asymmetries in the halos/shells; and evidence for enhanced low excitation emission at the leading edges of the shells, presumed to represent regions of shock interaction with the ISM.

Such qualitative analyses of the halos has subsequently been confirmed through though a broad range of analytical studies (Smith 1976; Borkowski et al. 1990; Soker & Dgani 1997; Dgani & Soker 1998; Dgani 1998, 2000) and hydrodynamical modelling (Soker et al. 1991; Brighenti & D'Ercole 1995a, b; Villaver et al. 2000, 2003; Szentgyorgyi et al. 2003; Pittard et al. 2005; Wareing et al. 2007a, b). These confirm the theoretical likelihood of CSs being shifted from the geometrical centres of the nebulae (e.g. Smith 1976); asymmetrical brightening and distortion of the shells (e.g. Brighenti & D'Ercole 1995a, b; Villaver et al. 2000, 2003; Szentgyorgyi et al. 2003); and ram-pressure stripping of the AGB envelopes, and the formation of cometary-type structures (Villaver et al. 2003; Wareing et al. 2007a, b). They also suggest that such interactions are likely to occur, and to have serious consequences from the very earliest stages of AGB mass-loss (e.g. Villaver et al. 2003; Wareing et al. 2007a, b). The modelling by Villaver et al. (2003), in which thermal pulses dominate variations in the progenitor mass-loss rates, show that a convex envelope is likely to form; a structure which will undergo differing phases of expansion as the equilibrium balance of pressure varies with time. However, Rayleigh-Taylor (RT) and Kelvin-Helmholtz (KH) instabilities are also likely to be important at the leading edges of the envelopes, and this may lead to the peeling-off of vortices, and large-scale transfer of AGB material into the nebular tails (Szentgyorgyi et al. 2003; Wareing et al. 2007a, b). It has even been suggested that a good fraction of the nebular mass may lie within these tails, thereby solving a disparity between observed nebular masses, and those that would be expected from initial-final mass analyses (Villaver et al. 2003, Wareing et al. 2007b). The roles of such instabilities has been studied by various authors, including Wareing et al. (2007a, b), whence they are shown to lead to oscillation of the equilibrium frontal position as material is transported towards the rear (ram-pressure stripping), and post-shock AGB material is replenished. It is also likely that the vortices may, in some cases, have an asymmetrical placement towards the rear, a situation for which we will find further evidence in the imaging to be presented below.

Most of the latter results fail to take account of the interstellar (IS) magnetic field, however, and assume simplified versions of the progenitor mass-loss history. Such failings may turn out to be critical. Thus, Tweedy, Martos & Noriega-Crespo (1995) pointed out that the stand-off radius for pressure equilibrium may depend strongly upon the strength of the IS magnetic field. A variety of papers by Dgani and Soker (see e.g. Soker & Dgani 1997; Soker & Zucker 1997; Dgani & Soker 1998; Dgani 1998, 2000) have expanded upon this theme, and show that the role of the IS magnetic field may be important in the formation of instabilities. A particular case of interest is where so-called "striping" occurs, whereby bands of emission are seen close to the leading or trailing edges of the PNe. Dgani & Soker (1998) argue that this is likely to arise where the flows are isothermal, and post-shock magnetic pressures exceed the thermal pressure of the gas. This leads to a suppression of instabilities along the direction of the magnetic field lines, and the growth of "rolls" or "stripes" in the direction of PN motion.

Dgani & Soker (1998) claim that such stripes have been observed in a surprising number of interacting PNe (they list 34 such sources), and show a propensity to occur at low galactic latitudes, where ISM densities are larger and isothermal flows more probable. Sources at higher galactic latitudes, by contrast, are expected to be ploughing through a much lower density ISM, and this leads to lower rates of post-shock cooling, higher thermal pressures, a dominance by shorter wavelength RT modes, and the suppression of elongated stripes.

These and other predictions by these authors need to be tested through detailed magneto-hydrodynamic modelling. We shall note however that such striping may be occurring in at least one of the sources to be considered below (NGC 3242).

Most observations of interacting shells have previously been obtained in the visual wavelength regime, although evidence for these processes has also been noted at infrared wavelengths as well (Meaburn et al. 2000; Zijlstra & Weinberger 2002). We shall present the first high resolution images of these shells in the mid-infrared, acquired using the Infrared Array Camera (IRAC; Fazio et al. 2004) mounted on the *Spitzer* Space Telescope (SST; Werner et al. 2004). These will confirm that the MIR structures for NGC 6772 and 6629 are similar to those noted in the visible, although 8.0 μm emission processes are significantly different. They will also be shown to be reminiscent of various of the modelling results cited above. The interaction of NGC 2438 with the ISM appears to be more modest, although our results show the outer structure of the source rather more clearly than is the case in the visible. It is suggested that irregularities in the outer shell may arise due to RT instabilities, expected to be important where shell densities decline to the critical ram-pressure equilibrium density $n_{CRIT}$ (Dgani 2000) - the nebular density at which the ISM ram pressure balances the PN thermal pressure. This latter parameter, in the absence of strong ISM magnetic fields, may be approximated by the relation by $n_{CRIT} \sim ((v_*+v_{exp})/c)^2 n_0$, where $n_0$ is the density of the ISM, c is the isothermal sound speed of the PN gas, $v_*$ is the velocity of the central star, and $v_{exp}$ of expansion of the nebular gas. It will also be noted that previous hydrodynamical modelling of the shell requires the central star luminosity to be relatively low – although it is pointed out that this is not, in fact, necessarily the case; it is conceivable that the central star luminosity is larger than has previously been deduced.

Finally, we shall note that there is prior evidence for a diffuse outer halo about the multi-polar outflow NGC 2440, although the structure of this envelope was by no means clearly defined. The present MIR results show that it is likely to represent an interacting envelope, and enables us to clarify the relation between the extended fainter shell and other (CO, $H_2$, HII) components of emission.

## 2. Observations

We shall be making use, in the following analysis, of data products deriving from three SST programs. The nebulae NGC 2438, NGC 6629, and NGC 6772 were targeted as part of program 68 (Studying Stellar Ejecta on the Large Scale using SIRTF-IRAC), and the observations took place respectively on 26/11/2004, 06/09/2004, and 05/10/2004. The raw results then went through the various stages of processing as described in the IRAC data handbook (available at http://ssc.spitzer.caltech.edu/irac/dh/iracdatahandbook3.0.pdf), the first of which results in the so-called Basic Calibrated Data (BCD). In this case, the raw observations are converted into an appropriately flux-calibrated image, and the primary instrumental defects are removed. A further stage of processing (post BCD) was then undertaken using a specific, conservative set of parameters, and included cosmetic corrections to the images – the removal of defects which are not based on well-established instrumental artifacts or detector physics. A so-called "pointing refinement" is also undertaken, whereby point sources within the fields are astrometrically matched to sources in the 2MASS catalogue, whilst mosaics are produced from the multiple AORs (Astronomical Observation Requests; in the case of Program 68, this involved combining 12 BCDs for each of the nebulae investigated here).

The present results all correspond to post-BCD products and are, as a result, relatively free from artefacts; are well calibrated in units of MJy sr$^{-1}$; and have reasonably flat emission backgrounds. An exception is NGC 6629, which has a weak central horizontal emission band at 8.0 μm caused by the bright central core. We have not removed this from the present results, since it has little impact upon the analysis or its interpretation. It does however need to be considered when obtaining profiles through the source.

The two other sources in this paper, NGC 2440 and NGC 3242, were observed through the respective programs 1052 ("IRAC observations of the butterfly-shaped nebula NGC 2899"; observation date 21/11/2003) and 30285 ("Spitzer observations of planetary nebulae 2"; observation date 29/12/2006). The results used here also correspond to post BCD data, and have similar 8.0 $\mu$m banding to that described for NGC 6629 above. For the former case (NGC 2440), the final image is a combination of 34 BCDs, whilst the NGC 3242 results represent a combination of 14 BCDs.

The observations were undertaken using the Infrared Array Camera (IRAC; Fazio et al. 2004), and employed filters having isophotal wavelengths (and bandwidths $\Delta\lambda$) of 3.550 $\mu$m ($\Delta\lambda$ = 0.75 $\mu$m), 4.493 $\mu$m ($\Delta\lambda$ = 1.9015 $\mu$m), 5.731 $\mu$m ($\Delta\lambda$ = 1.425 $\mu$m) and 7.872 $\mu$m ($\Delta\lambda$ = 2.905 $\mu$m). The normal spatial resolution for this instrument varies between ~1.7 and ~2 arcsec (Fazio et al. 2004), and is reasonably similar in all of the bands, although there is a stronger diffraction halo at 8 $\mu$m than in the other IRAC bands. This leads to differences between the point source functions (PSFs) at ~0.1 peak flux.

We have used these data to produce colour coded combined images of NGC 2440 and NGC 3242, where 3.6 $\mu$m is represented as blue, 4.5 $\mu$m as green, and 8.0 $\mu$m is indicated by red. The mapping of NGC 3242 also been processed using unsharp masking techniques, whereby a blurred or "unsharp" positive of the original image is combined with the negative. This leads to a significant enhancement of higher spatial frequency components, and an apparent "sharpening" of the image (see e.g. Levi 1974). Profiles through these sources have also been produced with the aim of evaluating the fall-off in surface brightness of the halo structures. This involved an initial correcting for the effects of background emission; a component which is present in all of the bands, but is particularly strong at 5.8 and 8.0 $\mu$m. The latter two bands are also prone to slight gradients in the background, of the order of 5 $10^{-4}$ MJy/sr/pix (although actual gradients vary depending upon the waveband and source under consideration, and the direction of the slice). The largest gradients were observed for NGC 6629. We have removed these trends by subtracting lineal ramps from the results – a procedure which is more than adequate given the limited sizes of the nebulae. Three of the sources (NGC 2440, NGC 3242, NGC 6629) also suffer from weak emission bands within the 8.0 $\mu$m images, as described above. The strength of these contaminants is relatively low (0.4-0.7 % of peak emission fluxes), however, and tends to be constant with x-axis displacement. We have chosen slices through the nebular nuclei which minimise the impact of these features, and such bands are likely to have zero-to-negligible influence upon our present results.

The results were subsequently processed so as to indicate the variation of 3.6$\mu$m/4.5$\mu$m, 5.8$\mu$m/4.5$\mu$m and 8.0$\mu$m/4.5$\mu$m ratios with distance from the nucleus. The rationale behind this is based on the fact that many PNe possess strong PAH emission bands at 3.3, 6.6, 7.7 and 8.6 $\mu$m, located in the 3.6, 5.8 and 8.0 $\mu$m IRAC bands. Furthermore these features, and their associated plateau components, show evidence for increased strength outside of the nebular cores, in the halo regions of interest to our present analysis. Given that the 4.5 $\mu$m band is usually dominated by bremsstrahlung continua and a variety of molecular and ionic transitions, it then follows that the variation of these ratios gives some insight into the importance of PAH emitting grains (see e.g. Phillips & Ramos-Larios 2008a, 2008b).

Some care must be taken in interpreting the flux ratio results, however. The problems with large aperture photometry are described in the IRAC data handbook, and relate in part to scattering in an epoxy layer between the detector and multiplexer (Cohen et al. 2007). This leads to the need for flux corrections as described in Table 5.7 of the handbook; corrections which are of maximum order 0.944 at 3.6 $\mu$m, 0.937 at 4.5 $\mu$m, 0.772 at 5.8 $\mu$m and 0.737 at 8.0 $\mu$m. However, the precise value of this correction also depends on the underlying surface brightness distribution of the source, and for objects with size ~several arcminutes it is counselled to use corrections which are somewhat smaller. The handbook concludes that "this remains one of the largest outstanding calibration problems of IRAC".

We have, in the face of these problems, chosen to leave the flux ratio profiles unchanged. The maximum correction factors for the 8.0μm/4.5μm and 5.8μm/4.5μm ratios are likely to be > 0.8, but less than unity, and ignoring this correction has little effect upon our interpretation of the results.

Finally, we shall be presenting an Infrared Space Observatory (ISO)[1] pipeline spectrum for NGC 2440. Sections of this data have previously been presented by Bernard Salas et al. (2002), and used to determine electron densities, temperatures, and elemental abundances within the source. The basic characteristics of these observations can be summarised as follows. The results were taken with the Short Wavelength Spectrometer (SWS), information concerning which can be found in de Grauuw et al. (1996). The data covers a region of size 14x20 arcsec$^2$ centred at RA(2000) = 07h 41m 55.4 s, Dec(2000) = -18° 12′31.8", and covers a spectral regime 2.38-12 μm. The exposure time was 1912 s, position angle was 23°, and the spectral resolution is 250. The spectrum appears not to be affected by fringes. We also provide a Spitzer pipeline spectra for NGC 6629. This was acquired as part of the program Deuterium Enrichment in PAHs, and was acquired on 05/02/2007. The aperture size was 3.6x57 arcsec$^2$ centred on RA(2000) = 18h 25m 42.67, Dec(2000) = -23d 12m 19.6s and with a slit position angle of 171.57°, whilst the spectral channels have a uniform width of 3.03 10$^{-2}$ μm.

## 3. MIR Observations of Planetary Nebulae

### 3(i) NGC 2438

Our 8.0 μm images of NGC 2438 are illustrated in Fig. 1, where it is apparent that the bright central nebula is surrounded by two extended halos. This image is similar to the corresponding results in [NII] + Hα (Corradi et al. 2003), although the outer easterly sectors of the halo are in this case somewhat better defined.

It is apparent that the peripheral limits of the outer halo show a scalloped appearance, whilst the inner halo is more circularly symmetric, and has a smooth and continuous outer rim. It is also apparent that the geometrical centre of the outer halo is displaced from that of the inner halo, as well as being displaced from the position of the central star. These two factors together suggest that (i) the source is moving towards the upper part of the panel, causing an easterly shift in the position of the outer structure, and (ii) that the outer parts of the halo are experiencing RT instabilities, expected to occur where the density of the shocked, cooler ISM is smaller than the post shock AGB halo (Dgani 2000). The time-scale for the formation of such instabilities is expected to be similar to that required for appreciable deceleration of the shell (Soker et al. 1991). Although the rim of the outer halo may arise as a result of direct interaction with the ISM, it is possible that it may develop as a result of the last helium shell flash on the AGB (e.g. Corradi et al. 2003). Both mechanisms would be likely to lead to similar, although not necessarily identical consequences. It is possible, in the case of shock interaction with the ISM, to envisage that the higher temperatures and densities in the AGB post-shock regime may lead to sputtering and destruction of small PAH type grains, for instance, and a modification in the MIR colour indices. Although the relevant zones are extremely faint in both the visible and MIR, we shall nevertheless make an investigation of such trends in our analysis below.

---

[1] *Based on observations with ISO, an ESA project with instruments funded by ESA Member States (especially the PI countries: France, Germany, the Netherlands and the United Kingdom) and with the participation of ISAS and NASA*

It is worth noting, in this respect, that the double halo in NGC 2438, and its associated rims, have be modelled in terms of AGB winds and the last thermal pulse of the progenitor star (for the outer halo), and in terms of a recombining shell and recombination front (for the inner halo) (Corradi et al. 2000). This latter type of structure has been described by Tylenda (1986) and Phillips (2000), and occurs where the stellar luminosity declines during later phases of evolution, and is insufficient to maintain a fully ionised nebular shell. The effective Stromgren radius declines, and recombination occurs.

In support of this argument, the authors cite evidence for central star temperatures of between $1.65 \cdot 10^5$ K and $1.14 \cdot 10^5$ K, and luminosities of between 120 and 570 $L_\odot$; values which although clearly uncertain, place the star on the declining branch of its HR evolutionary track.

There is, however, a problem with such a suggestion, and it is by no means certain that the analysis of these authors can be regarded as cut and dried. The first point to note is that Rauch et al. (1999), in determining the central star luminosity of 570 $L_\odot$, noted that the luminosity required to maintain the ionised shell is very much greater (of order 4.9 10$^3$ L$_\odot$). They specifically draw attention to this disparity, and consider it as a problem which requires to be resolved.

In addition, we note that were one adopts the central star visual magnitude m$_v$ = 18.0 mag of Rauch et al. (1999); uses their quoted extinction c = 0.67; employs the much lower statistical distance of d =1.20 kpc of Phillips (2004) (as compared to the value d = 4.3 kpc quoted by Rauch et al. 1999); and employs a mean HeII Zanstra temperature T$_Z$(HeII) = 3.43 10$^5$ K for the stellar effective temperature (Phillips 2003a), then using an equation

$$\frac{L}{L_{SUN}} = 4.74 \, 10^{10} \left[\frac{d}{kpc}\right]^2 \left[\frac{\lambda}{\mu m}\right]^5 \left[\frac{c_\lambda}{erg \; cm^{-2} s^{-1} A^{-1}}\right] \left[\frac{T_{eff}}{10^3 K}\right]^4$$

$$\times 10^{-0.4(m_\lambda - A_\lambda)} \left[\exp\left\{\frac{14.388310}{(\lambda/\mu m)(T_{eff}/10^3 K)}\right\} - 1\right] \quad \ldots\ldots (1)$$

deriving from Gathier and Pottasch (1989), where d is the source distance, m$_\lambda$ and A$_\lambda$ are the apparent magnitude and extinction of the central star, and c$_\lambda$ is derived from the flux calibration of α Lyrae by Tüg et al. (1977) (c$_\lambda$ = 3.59 10$^{-9}$ erg cm$^{-2}$ s$^{-1}$ at V), then one determines a luminosity L$_*$ ≅ 3.16 10$^3$ L$_\odot$. Alternatively, where one uses the statistical distance d = 2.09 kpc of Zhang (1995) then one obtains L$_*$ ≅ 9.6 10$^3$ L$_\odot$.

Such values are of course still highly uncertain; the luminosities would be very much larger where one uses the quoted distance of Rauch et al. (1999), but somewhat less where one employs the extinction coefficients of Tylenda et al. (~ 0.25-0.35). Similarly, the equation of Gathier & Pottasch (1989) assumes that the central star continuum can be modelled using a blackbody relation. This is, of course, far from being the case, particularly where stellar temperatures are as high as they are likely to be for this particular source. It is therefore follows that L$_*$ may, yet again, have been somewhat understated.

Nevertheless, and despite these caveats, it is clear that the luminosity $L_*$ is likely to be greater than that supposed by Rauch et al. (1999), and more comparable to the value that Rauch et al. state as necessary for ionisation of the nebular envelope.

There is also further evidence for high CS luminosities in this particular source. Although accurate bolometric corrections (B.C.) are unavailable for stars having such high effective temperatures (see e.g. the discussion of Phillips 2005), the relation B.C. = 27.76 - 6.87log($T_{eff}$) (Phillips 2005) based upon the data of Méndez et al. (1992) is probably reasonably applicable to the present source. Using this relation, and again employing the values $m_v$ and c of Rauch et al. (1999), one determines a CS luminosity $L_* \sim 3.41 \; 10^3 \; L_\odot$ for the statistical distance of Phillips (2004), or a value ~ 3.0 times greater for the distance of Zhang (1995). These values are reasonably similar to those deduced from the analysis above.

These combined luminosities and temperatures therefore place the star close to the turn-over point of its evolutionary track, providing that it has a mass $M_{CS} \sim 0.9 \; M_\odot$, and progenitor mass $M_{PG} \sim 5.0 \; M_\odot$ (Vassiliadi & Wood, 1994). It would, under these circumstances, be implausible to expect a fully developed recombination halo of type hypothesised by Corradi et al. (2000).

Finally, profiles through the shell are shown in Fig. 2, where we indicate trends in all four of the photometric channels. The inner rim, and one portion of the outer rim are clearly to be discerned, and show a tendency for emission to increase with increasing MIR wavelength. Flux ratios for other portions of halo are extremely difficult to determine, not least because flux levels are mostly lower; the uncertainties in background subtraction become even more severe; and integrated fluxes at 3.6 and 4.5 μm are affected by the population of weak background stars. Insofar as can be determined, however, flux ratios for the halo and rims appear to be qualitatively similar. We determine halo values 0.58, 1.65 and 2.75, and rim measures of 0.53, 2.09, and 3.0 for the flux ratios 3.6μm/4.5μm, 5.8μm/4.5μm, and 8.0μm/4.5μm, after subtracting underlying halo fluxes from the rim measures of emission. These values are equal within the uncertainties.

**3(ii) NGC 2440**

Our MIR image of this source is illustrated in the upper panel of Fig. 3, where the colour coding is as indicated in Sect. 2. Two aspects of the source are immediately apparent. First, there is a high level of longer wave emission within the core of the source, indicated by the red-white colouration within ~ 8 arcsec of the central star. This also applies to two external ansae located ~ 30 arcsec to the left and right of the image (i.e. to the NE and SW of the centre), representing the limits of a bipolar kinematic structure investigated by Lopez et al. (1998) and others (e.g. Cuesta & Phillips, 2000; Wang et al. 2008).

There is also evidence for a roughly circular external halo which appears to show evidence for interaction with the ISM, and is sharply defined to the NE (the presumed direction of motion) and somewhat irregular and diffuse to the SW. Note also the much fainter emission to the upper right-hand side of the panel, and a central band of 8.0 $\mu$m emission described in Sect. 2. A comparison with the model examples illustrated in Fig. 4 suggests a structure which is comparable to that of Villaver et al. (2003), with perhaps some evidence for ram-pressure stripped material to the rear of the source, not dissimilar to the vortex-like structures modelled by Wareing et a. (2007b). The weak emission to the upper right is not balanced by any corresponding emission to the lower right, and is therefore reminiscent of the asymmetric tail noted in the upper central panel of Fig. 4.

The outer halo has been observed in the optical by Reay et al. (1988), and is also apparent in Digital Sky Survey images of the source (see e.g. Wang et al. 2008), although the visual results are more diffuse and less well defined. We shall discuss, below, how this structure relates to other elements of emission.

The lower left-hand panel of Fig. 3, for instance, shows a superposition of the ionised envelope observed by the HST upon our present 8.0 $\mu$m results (shown in black & white). In this case blue corresponds to the blue continuum, combined with HeII and [OIII] (taken with the F439W, F469N and F502N WFPC 2 (Wide-Field Planetary Camera 2) filters); green corresponds to the visual continuum and H$\alpha$ (F555W and F656N filters); and red to the R continuum and [NII] (the F675W amd F658N filters). It is fascinating to see that the outer limits of the hour-glass structure closely correspond to those of the 8.0 $\mu$m envelope – and indeed, that there is some extension of ionised emission outside of the hour-glass structure, and along the rim of the MIR shell, resulting in what appear to be delicate filaments attached to the easterly lobe. It is therefore clear that portions of the outer rim of the 8.0 $\mu$m halo are likely to be photoionised, whilst other parts may be shock ionised through interaction with the ISM.

A further interesting feature of the relation between the HII region and halo is that although much of the halo may be neutral (see our further comments below), the limits of the inner hourglass structure are congruent with those of the halo – there is no evidence that higher thermal pressures within the HII region are causing it to puncture the outer skin of the halo. This may imply that the halo is also largely ionised, albeit that it contains neutral inclusions, and/or that the hourglass structure is in pressure equilibrium with the shocked ISM.

A final panel to the lower right-hand side of Fig. 3 shows the relation between the 8.0 $\mu$m halo and molecular emission. We have in this case represented a map of $H_2$ S(1) emission (indicated as green) and CO J = 3-2 at the systematic velocity of $V_{LSR}$ = 43 km s$^{-1}$ (blue contours), both of which derive from Wang et al. (2008). Previous $H_2$ measures of the source have been published by Reay et al. (1988), Latter et al. (1995) and Kastner et al. (1996), whilst lower frequency CO measures are available in Dayal & Bieging (1996) and Huggins et al. (1996).

It is immediately apparent that $H_2$ emission is found in most parts of the MIR halo, and has a structure which is closely similar to that observed at 8 $\mu$m. The only exceptions occur in the upper right and lower-left hand portions of the 8.0 $\mu$m shell, where the $H_2$ emission appears mostly to be absent. It is perhaps no coincidence that these are also the sectors where the hourglass structure is to be found, suggesting that local ionisation is resulting in destruction of the molecules.

It is known that $H_2$ emission is present in the halo regions of many PNe (Kastner et al. 1994; Latter & Hora, 1997; Ramos-Larios, Guerrero & Miranda, 2008), and has potentially strong transitions within the MIR (Neufeld & Yuan 2008). It is therefore entirely possible that the MIR emission in NGC 2440 is dominated by shock or fluorescently excited $H_2$ emission.

This is not, however, the only possible explanation, or even the most likely. The ISO spectrum of the source has been analysed by Bernard Salas et al. (2002), and we show a segment of this data in Fig. 5. This corresponds to a region of size 14x20 arcsec$^2$ centred upon the nucleus of the source. Not all of the lines are clearly identifiable, although we have flagged a large fraction of the transitions in the lower part of the figure. None of these are identifiable with H$_2$, even though it appears that such emission is likely to be strong within small distances of the central star (e.g. Reay et al. 1988; Latter et al. 1995; Kastner et al. 1996; Wang et al. 2008).

The line at 9.25 $\mu$m may be spurious, and result from responsivity corrections (see e.g. the SWS Instrument Data Users Manual at http://www.ipac.caltech.edu/iso/sws/idum/sws_idum.html).

Two things are apparent from the spectrum. The first is that there is little continuum emission, apart from a possible weak component between 5.8 and 7.0 $\mu$m. Most of the MIR emission in the nucleus and halo is likely to arise from ionic forbidden and permitted line transitions. The second point worth noting is that we are observing a broad range of excitations within the source, from very high ionisation potential (IP) ions such as Ne$^{5+}$, through to lesser IP ions such as H$^+$ and Ar$^+$. It is therefore possible that the red colouration in the nucleus (Fig. 3,upper panel) is a consequence of strong higher excitation lines such as [NeVI], and that the outer halo of the source has much lower levels of excitation, and stronger emission in the lower wavelength bands.

If this is the case, however, then it is unlikely to explain the red colour of the ansae to the right and left of the Spitzer image. In this case, we note that the ansae and their vicinities may be relatively dusty (see e.g. the comments of Lopez et al. 1998), however, and it is conceivable that 5.6 and 8.0 $\mu$m emission is enhanced as a result of PAH emission bands, the wings of the 11 $\mu$m SiC emission feature, and/or broader grain continuum components of emission. This, if true, would be consistent with the mean abundance ratios C/O for this source, which imply a slight excess of C with respect to O (Phillips 2003b).

There is also clear evidence that the ansae and the nucleus are associated with strong molecular components of emission, as is testified by the CO J = 3-2 contours in the lower right-hand panel of Fig. 3, and enhanced levels of H2 S(1) emission (more clearly seen in the imaging of Wang et al. (2008)).

It is therefore apparent that the outer MIR halo is partially ionised at the NE leading edge, and is partially co-spatial with a $H_2$ emitting shell, although the emission mechanisms responsible for this structure are from being clearly defined. The ISO spectrum of the source, and the lack of complete equivalence between the $H_2$ and MIR emitting shells, suggest that the role of any $H_2$ transitions is likely to be modest.

Profiles through the source are illustrated in Fig. 6, where we have taken a diagonal slice designed to show the nature of emission towards the leading and trailing edges of the halo. We have also shown the corresponding variation in flux ratios in the lower panel of this figure. Several things are apparent from these graphs. In the first place, it is clear that the inner portions of the source have large 8.0$\mu$m/4.5$\mu$m flux ratios. These may, as noted above, reflect the tendency for [NeVI] $\lambda$7.642 $\mu$m emission to be preferentially concentrated towards the nucleus. This ratio declines however towards larger distances from the nucleus, where it becomes comparable to the corresponding 5.8$\mu$m/4.5$\mu$m ratios. We also note that the left-hand rim of the source corresponds to a regime of strongly reduced 8.0$\mu$m/4.5$\mu$m ratios, and a somewhat lesser reduction in 5.6 $\mu$m/4.5 $\mu$m – suggesting that whatever is responsible for emission in the longer wave bands is partially suppressed within the rim. This does not, however, occur within the right-hand rim (i.e. that located to the south of the central star).

Finally, the profiles in the upper part of the figure show the rapid decline in shell surface brightness with increasing distance from the central star, terminating in the rim enhancements at relative positions + 42.3 arcsec and –29.8 arcsec. This is not, however, the end of the story. It appears that the fall-off in MIR emission may also extend beyond the rim, as is more clearly visible to the left-hand side of the graph. Whilst the 3.6 and 4.5 $\mu$m emission falls-off very steeply for RP < - 30 arcsec, the fall-off at 5.8 and 8.0 $\mu$m is very much shallower; although it is still steeper than the decline noted in the halo of the source itself. There is therefore evidence for continuing emission from beyond the limits of the rims.

Some caution should be exercised in interpreting these results, however, given the effects of longer-wave scattering in the IRAC camera (see Sect. 2). If such trends prove to be real, however, then it might suggest that the interaction interface with the ISM lies at even larger negative relative positions, and well outside of the bright-rimmed structure apparent in Fig. 3.

**3(iii) NGC 3242**

Our 8.0 $\mu$m image of NGC 3242 is illustrated in Fig. 1, and shows both the inner ringed halo described by Phillips et al. (2009) and a more diffuse, outer structure which is apparently subject to severe fragmentation and instability. We note, in this respect, that the halo is rather similar to the structures predicted by Dgani & Soker (1998), whereby RT instabilities at the leading edge result in an opening of the structure, and flow of ISM material into the inner portions of the shell. An illustration of this mechanism is provided in Fig. 4 (lower left-hand panel), although it should be admitted that the inner ringed halo in NGC 3242 shows little evidence for disruption by the ISM. Similarly, the model appears to show only a single opening as a consequence of RT instabilities, in stark contrast to the shredded appearance noted in Fig. 1. It has been noted however that full three dimensional modelling of these outflows is likely to "result in many more RT "fingers" (or "tongues")" (Dgani & Soker 1998).

On this basis, therefore, it seems possible to hypothesise that the nebula is moving towards the upper left hand side of the panel (~north), and that RT instabilities at the ISM/nebular interface are leading to compression and fragmentation of the nebula, pushing this part of the shell towards the central star. The southerly portions of the shell, by contrast, are more diffuse and extended, and may contain ram-pressure stripped portions of the AGB envelope.

So far so good – but is there any other evidence to support such a presumption?

An indication that there is such evidence is provided in Fig. 7, where we have combined a deep H$\alpha$ image due to Corradi et al. (2003) (coloured blue) with a Spitzer image of the source. This latter figure has been processed using unsharp masking techniques so as to emphasise finer details of the nebular structure, whilst the colour coding is as indicated in Sect. 2. We also show a train of 60 μm emission deriving from a "maximum correlation" analysis of the IRAS results, taken from a map provided by Meaburn et al. (2000). Several things of interest are apparent in this figure. In the first place, the direction of motion which we adduced from the halo of the source, and which is indicated by the red arrow to the upper left-hand side of Fig. 7, is diametrically opposed to the trail of 60 μm emitting material noted by Meaburn et al. (2000). This suggests that this latter feature corresponds to ram-pressure stripped gas and grains; material which was originally located within the AGB outflow envelope. A second feature of interest is the presence of a couple of bow-shock type waves to the upper left-hand side of the source – features which may represent examples of the magnetically induced stripes described by Dgani & Soker (1998), Dgani (1998, 2000) and Soker & Zucker (1997). A panel schematically illustrating the formation of such stripes or "rolls" is illustrated in the lower right-hand panel of Fig. 4. Dgani (1998, 2000) and Dgani & Soker (1998) find that the separation of the stripes $\Delta Z$ is of order ~ $Rv_{A0}\cos\alpha_0/(2^{0.5}v_*)$, where $v_{A0}$ is the Alfven speed of the ISM, R is the radius of the nebula, $\alpha_0$ is the angle between the pre-shock magnetic field and the shock front, and $v_*$ is the velocity of the central star. It has therefore been proposed that the separation of the stripes may enable us to probe the strength of the IS magnetic field.

Too many uncertainties apply to the present case to enable us to sensibly constrain the value of B. However, it is clear that the stripes are located precisely where one might expect them to be – just outside of the primary shell, and in the direction of motion to the upper left-hand side. It is worthwhile noting, in this respect, that although the direction of the ISM magnetic field may be strongly non-perpendicular to the direction of motion, the perpendicular vector through the "stripes" is expected to be closely parallel to the CS velocity vector $\vec{v}_*$ (Dgani 1998). It follows that most of the information concerning the orientation of the magnetic fields is lost, and cannot be ascertained from the direction of the "stripes".

A complicating aspect of this analysis, persuasive though it may seem, is that the Galaxy Evolution Explorer (GALEX) UV/optical images of the source show a somewhat differing morphology (see e.g. images published at http://www.galex.caltech.edu/media/images/glx2009-03r_img01_Sm.jpg). Although stripes still appear to be present, and are located in the same position as those in Fig. 7, they now have a linear structure, and appear to be connected to filaments at larger distances from the source. They appear, in brief, to be little more than structural features of the larger enveloping cloud. It is somewhat perplexing to note such marked changes in morphology between the UV and H$\alpha$, although it may suggest that we are dealing a chance superposition of unrelated features. Alternatively, it may be possible that the proposed magnetically induced stripes are unstable over very short periods of time, and that differences between the GALEX and H$\alpha$ images are indicative of extraordinarily interesting large scale variations.

**3(iv) NGC 6629**

An 8 $\mu$m image of NGC 6629 is illustrated in Fig. 1, where it is clear that the structure is in certain ways similar that that for NGC 2440 – although the evidence for shock deformation of the halo is even stronger. In this case, one presumes that the central star and PN are moving towards the N, and that this leads to compression of the shell, a bright rim-like structure, and arms of emission similar to those noted in the modelling of Villaver et al. (2003) (Fig. 4). There is also again evidence for emission to the upper left-hand side of the source, possibly representing ram-pressure stripped material, and the presence of vortex-type instabilities to the rear (viz. the models of Wareing et al. (2007a, b) in Fig. 4). It would be fascinating to obtain even deeper imaging of the source with the new generations of space-borne infrared telescopes.

Much can nevertheless still be done using ground-based instrumentation alone, since a superposition of the present 8.0 $\mu$m image upon the H$\alpha$ imaging of Corradi et al. (2003) reveals that both of the shells are well nigh indistinguishable – both show exactly the same halo structure, relative halo intensities, and trailing elements of ram-stripped material. One might therefore conclude that both of the images are associated with similar emission mechanisms.

This is, however, far from being the case. Profiles through the source are illustrated in Fig. 8, where it is clear that fluxes are similar in the 3.6, 4.5 and 5.8 μm photometric channels. However, emission at 8.0 μm is clearly very much greater, and implies that there are unusually strong transitions within the 8.0 μm band, and/or that there is appreciable emission deriving from grain emission bands and/or continua.

That it is likely to be the latter explanation which is the more correct is indicated in Fig. 5, where we show the results of Spitzer spectroscopy between 5.16 and 8.6 μm. It is clear from this that whilst there is likely to be weak emission associated with the 6.2 μm PAH emission band, the three lowest wavelength IRAC photometry channels are likely dominated by forbidden and permitted line transitions. A smooth, strong, and increasing grain continuum only really kicks-in within the 8.0 μm channel, where there is also a very broad feature centred at 7.48 μm, presumably associated with PAH components of emission. Having said this, however, we note that the peak wavelength of this emission appears to bear little relation to normal PAH transitions within this regime (~6.9, 7.6, 7.8 & 8.6 mm, corresponding to differing C-C and C-H bending and stretching modes; see e.g. Tielens 2005).

Much depends in this case, however, upon whether the PAH grains are neutral or ionised; whether they are small or in the form of much larger PAH grain clusters; and what the excitation temperature of the bands might be. All of these factors can affect relative band strengths, and band emission profiles.

It therefore seems that the increase in 8.0 μm emission noted in Fig. 8 can be attributed to strong continuum and band emission due to warm ($\approx$ 400 K) nebular grains.

Apart from this, we note that the structures of the profiles are again similar to those noted for NGC 2440. There is again the steep fall-off in nebular emission beyond the nebular rim – less steep at 5.8 and 8.0 μm than is the case for 3.6 and 4.5 μm (upper panel of Fig. 8). There is also the dip in 8.0 μm emission at the positions of the rim (lower panel of Fig. 8). This latter behaviour, in the present source, would clearly be indicative of destruction of certain of the grain species, perhaps involving the depletion of PAH grains within the inhospitable environment of the shock. If this is the case, however, then it may be necessary to argue for a similar component of grain emission in NGC 2440 as well, since this shows similar dips in the strength of the 8.0 $\mu$m fluxes.

A difference with NGC 2440, on the other hand, is that 8.0 $\mu$m emission appears to increase within the nebular halo regime – that is, between the positions of the rims and main interior ionised shell – and this leads to the peaks in 8.0$\mu$m/4.5$\mu$m flux ratios in the ranges -15 < RP < -7 arcsec,  and 7 < RP < 12 arcsec.

### 3(v) NGC 6772

Our 8.0 $\mu$m image of the source is illustrated in Fig. 1, and represents an almost carbon copy structure to that noted in the visible (Corradi et al. 2003). Note in particular the bright rimmed region to the upper right hand side (the SW), presumably corresponding to shocked and fragmented AGB material; the fan-like distribution of material extending ~3 arcmin to the NE; and evidence for ray-like enhancements in the interior of the halo. It may very well be that we are observing RT disruption of the outer structure, not dissimilar to that illustrated in the lower left-hand panel of Fig. 4, although the level of frontal instability is greater than in the schematic modelling of Dgani & Soker (1998). This difference may become less apparent when full three dimensional simulations are considered, as noted in our analysis in Sect. 3(iii). It is also conceivable that the ISM is flowing into the inner portions of the halo, so that the original AGB envelope now no longer exists, but has been swept into trailing sectors of the shell. This cannot be directly deduced from our current imaging, but is a plausible consequence of the modelling of Dgani & Soker (1998).

A profile across the nebula is illustrated in Fig. 9, where we also illustrate MIR flux ratios, and highlight the rim structure of the source. Some care must be taken in interpreting these latter features, since the weak 3.6 and 4.5 $\mu$m emission is distorted by underyling stellar continua. This is responsible for the narrow peaks in emission between relative positions -147 and -153 arcsec, and 76 and 80 arcsec. Putting these contaminants to one side, however, it is plain, from the overlap of the MIR images, that the structures of the rims are similar in all of the IRAC wavelengths. It is also evident that there is an increase in flux levels with increasing photometric wavelength, similar to that noted in other portions of the halo in this source.

The 8.0μm/4.5μm, 5.8μm/4.5μm and 3.6μm/4.5μm flux ratios are seen to increase with increasing distance from the nucleus (lower panel of Fig. 9), a trend which has previously been attributed to increasing PAH emission within the halos of PNe (e.g. Phillips & Ramos-Larios 2008a, b; Ramos-Larios & Phillips 2008), but may also be influenced by shock or fluorescently excited $H_2$ transitions. Some evidence that the latter mechanism may be important in this particular source is provided through $H_2$ S(1) imaging at 2.12 μm (e.g. Webster et al. 1988; Kastner et al. 1994). There is also evidence for strong CO emission associated with the central elliptical ring (Bachiller et al. 1993). It is therefore clear that molecular material interlaces the inner and outer sectors of the shell.

## 4. Conclusions

We have presented mid-infrared imaging of planetary nebulae which appear to be interacting with the ISM. The sources show evidence for asymmetries in their halos, bright and narrowly defined leading edges, and displacement of the central stars from the geometrical centres of the outer envelopes. Certain of the nebulae have already been identified as interacting PNe on the basis of prior imaging in the visual wavelength regime, although the present images show features in these sources which have not previously been observed. It is clear for instance that there is scalloping of the outer halo of NGC 2438, probably arising due to Rayleigh-Taylor instabilities, with the largest deformations occurring at the easterly limits of the shell. Shifting of the central star with respect to the geometrical centre of the halo may suggest motion of the source towards the south. Although the dual-halo structure of the source may arise as a result of declining central star luminosities (Corradi et al. 2000), it is noted that CS luminosities may be higher than has previously been supposed. This, if true, may invalidate some of the assumptions in the Corradi et al. analysis.

A more extreme form of interaction appears to be present in NGC 6772, where the source appears to be moving towards the SW, and shows evidence for a fragmented frontal structure, as well as a considerable extension to the NE. It is possible that the source is experiencing instabilities similar to those described by Dgani & Soker (1998), wherein the ISM is flowing into the inner halo through spaces in the disrupted frontal structure.

NGC 3242 also shows a similarly fragmented structure to the NE side of the outer halo, suggesting very high levels of RT instability. The presumed direction of motion is also confirmed through previous observations of the source, which show the presence of an extended 60 $\mu$m tail of ram-pressure stripped material, and evidence for "striping"" in ionised material ahead of the source – a series of bands which may arise through RT instabilities in the magnetised ISM.

Although a visual halo has previously been observed in NGC 2440, its structure and characteristics were very poorly defined. The present observations suggest that it is again shock interacting with the ISM, and that the nebula is moving towards the NE. There may be evidence for ram-pressure stripped material to the rear (SW) of the source, perhaps arising through similar vortex-type instabilities to those described by Wareing et al. (2007a, b) and Szentgyorgyi et al. (2003). The limits of the hour-glass shaped ionised structure also correspond to those of the MIR halo, and there is some evidence that portions of the rim structure outside of the hour-glass are ionised as well – perhaps as a result of shock interaction with the ISM. Such a sharply defined temperature/ionisation structure is consistent with models of shell interaction (e.g. Wareing et al. 2007b), and may represent the first evidence for this phenomenon in PNe.

It is interesting to see that the ansae of one of the bipolar flows appear to reside at the limits of the halo, and that the structure of the halo is similar to that of the extended $H_2$ emission. Although it is unclear what emission agents are contributing to emission within the halo, MIR spectra of the nucleus appear to indicate an absence of strong continuum emission – whether it be attributable to dust or gaseous thermal contributions. It is also unlikely that the contribution of $H_2$ transitions can be very large, although it is apparent that such emission is present towards the nucleus of the source (the regime of the ISO spectrum) as well as in the more extended halo of the nebula. The largest contribution to MIR fluxes appears to be associated with permitted and forbidden line transitions, with the [NeVI] $\lambda$ 7.642 $\mu$m line being particularly strong within the 8.0 $\mu$m band. Variations in the relative strengths of these lines are probably responsible for changes of colour within the MIR map. It is possible however that the strong longer wave fluxes associated with the two principal ansae arise from locally high concentrations of PAH emitting grains. The ansae also appear to be repositories of particularly high concentrations of molecular gas, as evinced through CO J = 3-2 mapping by Wang et al. (2008).

Finally, NGC 6629 appears in some ways similar to NGC 2440, although the evidence for ISM/nebular interaction is perhaps even more severe. We note evidence for possible ram-pressure stripped material to the rear of the source, as was also found to be the case for NGC 2240. The structure at 8.0 $\mu$m is very similar to that noted in H$\alpha$, even though emission at these wavelengths is enhanced through grain continuum and band features. This suggests that the grains and gas are very closely intermixed. Something of an exception to this rule is found in the region of presumed shock interaction – the bright rim enhancement at the limits of the source. In this case, the relative ratios 8.0$\mu$m/4.5$\mu$m and (to a lesser degree) 5.8$\mu$m/4.5$\mu$m appear to dip below values in adjacent regions of the shell. This may imply destruction of the grains in the region of shock interaction.

This latter behaviour is also found in NGC 2440, although it is not entirely clear whether it is also present in the other sources considered here. The evidence, as it at present stands, suggests that these dips are a characteristic of two of the sources alone.

**Acknowledgements**

This work is based, in part, on observations made with the Spitzer Space Telescope, which is operated by the Jet Propulsion Laboratory, California Institute of Technology under a contract with NASA. GRL acknowledges support from CONACyT (Mexico) grant 93172.

**Figure Captions**

**Figure 1**

8.0 μm imaging for four of the five PNe investigated in this study. NGC 2438 has a double halo structure which has been attributed to declining luminosity in the central star. The outer halo shows evidence for RT instabilities, whilst the geometrical centre of this halo is displaced from the position of the central star (indicated both here and in the other panels by a central black circle). NGC 3242 shows strong disruption of the shell to the left-hand side, and interior rings which have been described by Phillips et al. (2009). NGC 6629 appears to be strongly compressed to the right (i.e. towards the north), and also shows evidence for blobs of emission to the upper left-hand side – a possible consequence of KH instabilities at the ISM/nebular interface, and ram-pressure stripping of AGB shell material. Note that the horizontal emission band represents an instrumental artefact. Finally, NGC 6772 appears to be strongly interacting to the upper right-hand side of the source (the SW), appears to the disrupted to the NE, and shows evidence for internal rays.

**Figure 2**

Profiles through the centre of NGC 2438, where levels of emission tend to be greater towards longer MIR wavelengths. Three portions of the rims are clearly apparent, and are flagged within the figure. The width of the slice is 6 pixels (equivalent to 7.2 arcsec), whilst its direction is indicated in the inserted figure.

**Figure 3**

Three representations of the planetary nebula NGC 2440. In the first of these (upper panel) we have illustrated a colour image of the source in the MIR, where 3.6 μm emission is represented as blue, 4.5 μm fluxes are green, and 8.0 μm emission is red. Note the presence of strong colour gradients between the core, halo and ansae, likely arising due to variations in permitted and PAH band emission components. It is suggested that the source may be moving to the NE, causing some compression of the shell to the left-hand side; the enhanced and narrow rim-like structure; and weaker components of emission to the upper right. These latter may arise due to KH instabilities at the leading edge of the source, and ram pressure stripping of AGB material. Note that the central red emission band is an artefact of the observations. The lower left-hand panel shows an HST image of the source taken from http://hubblesite.org/newscenter/archive/releases/2007/2007/09/& usg, superimposed upon a B&W image of the 8.0 $\mu$m structure. It can be seen that the outer limits of the hourglass structure are coincident with those of the halo. There is also evidence for separate ionisation of the upper left-hand portion of the halo rim. The colour coding of this figure is described within the text. Finally, the lower right-hand image shows J = 3-2 CO contours (blue) and extended $H_2$ S(1) emission (coloured green), both adapted from the work of Wang et al. (2008). These images are superimposed, yet again, upon a B&W 8.0 $\mu$m image of the source. Note the close similarity between the distributions of $H_2$ S(1) and 8.0 $\mu$m emission, although they are not completely identical. $H_2$ emission decreases to the lower left- and upper right-hand portions of the 8.0 $\mu$m structure, in locations where ionised emission (lower-left panel) is preferentially strong.

Finally, it is clear that CO emission is particularly strong in nucleus, and in the two primary ansae.

**Figure 4**

The results of schematic analyses, and hydrodynamical modelling of interacting PNe, taken from previously published results. The model of Villaver et al. (2003) (upper left-hand model) shows the results expected for interaction of PNe shells having strongly variable AGB mass loss rates. Those of Wareing et al. (2007a) (upper central panel) and Wareing et al. (2007b) (upper right hand panel) show the interactions expected for AGB envelopes, and at later stages of evolution where the PN shell has formed (inner bright ring). Both show evidence for turbulent wakes arising from K-H instabilities at the shock front. Note the asymmetries present in some of the vortex instabilities, which lead to corresponding asymmetries in the nebular wakes. Finally, we show two examples of analytical studies taken from Dgani & Soker (1998) where (lower left hand panel) it is suggested that RT instabilities may lead to disruption of the fronts, and the flow of ISM material into the inner portions of the halos. The lower right-hand panel shows the formation of "rolls" or "stripes" within the magnetised ISM – features which may arise ahead of, or to the rear of the primary ionised shells.

**Figure 5**

Spectra for NGC 6629 and NGC 2440, corresponding respectively to Spitzer and ISO pipeline products. We have identified the principal line features, and the ranges corresponding to the respective IRAC photometric bandpasses. Two broad emission features in NGC 6629 are identified as likely PAH emission bands.

**Figure 6**

The logarithmic variation of MIR surface brightness through NGC 2440 (top panel), where we show results for all four IRAC photometric channels. The corresponding variation in flux ratios is shown below. The rim features are again clearly in evidence and, in the case of the left-hand rim (at RP = -30 arcsec), appear to be associated with a dip in relative 8.0 $\mu$m flux levels. The width of the slice is 7 pixels.

**Figure 7**

The environment of NGC 3242, where we show a deep H$\alpha$ image taken by Corradi et al. (2000) (represented as blue); an MIR colour image of the interior portions of the source (where 3.6 $\mu$m is represented as blue, 4.5 $\mu$m emission by green, and the 8.0 $\mu$m results are red), processed using unsharp masking techniques; and evidence for a trail of 60 $\mu$m emission indicated by the red contours (adapted from Meaburn et al. (2000)). Notice the evidence for waves or "rolls" of emission in the presumed direction of motion (indicated by the red arrow), which may be attributable to RT instabilities in the magnetised ISM. The red diagonal band in the MIR image is an artefact of the observations.

**Figure 8**

As for Fig. 6, but for the case of NGC 6629. The profiles appear qualitatively similar to those for NGC 2440, showing as they do clear rim decreases in the relative strength of 8.0 $\mu$m emission. The width of the slice is 2.3 pixels.

**Figure 9**

As for Fig. 6, but for the case of NGC 6772. Emission close to the rims is highlighted in the two smaller inserted graphs. Although it is difficult to assess trends in flux ratio in the outer portions of the halo, it seems unlikely that the rim structure has flux characteristics which are in any way exceptional. Note the increase in flux ratios with increasing distance from the nucleus, a characteristic which has also been noted in other PNe (e.g. Phillips & Ramos-Larios 2008a, b; Ramos-Larios & Phillips 2008). The width of the slice is 6 pixels.

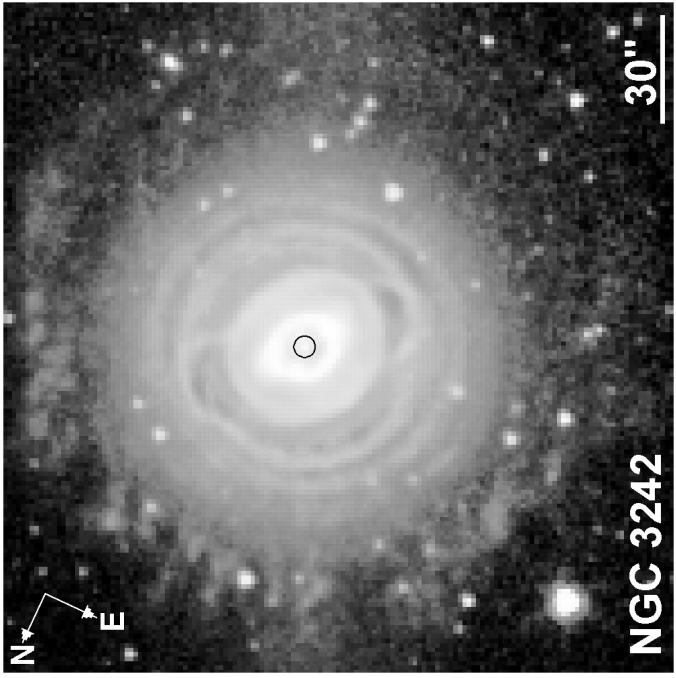
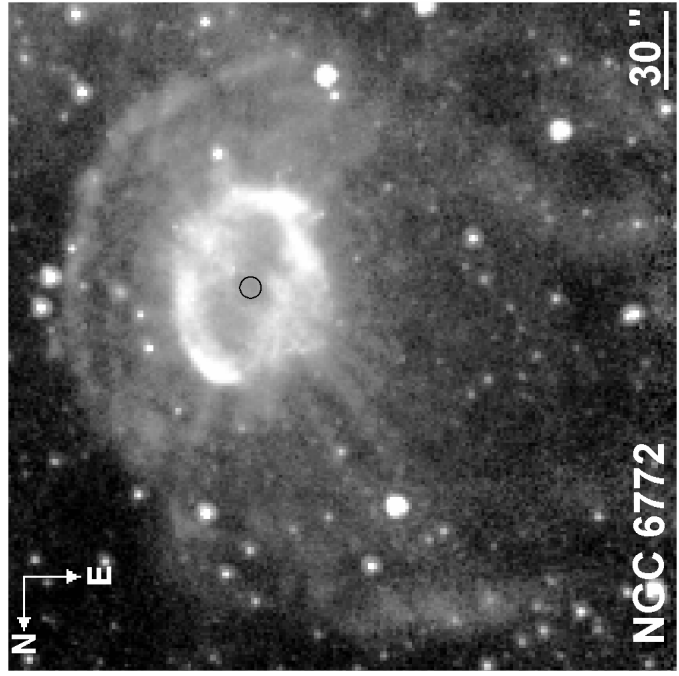
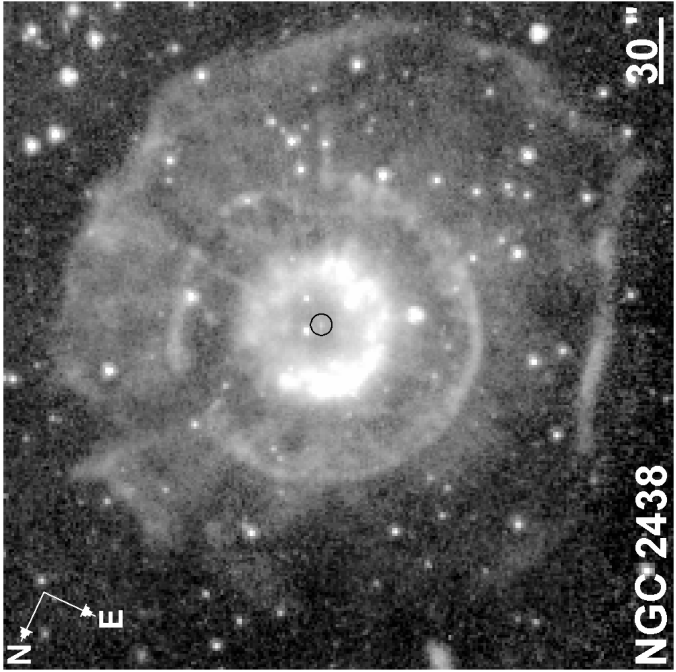
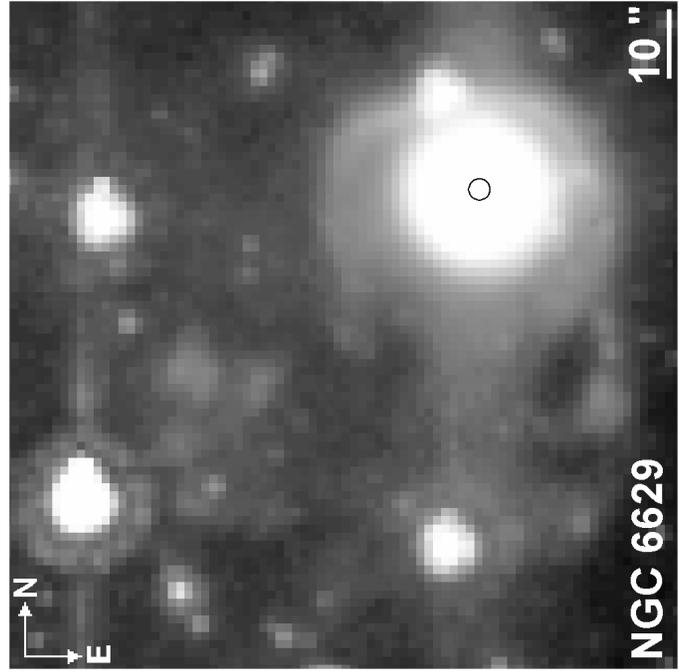

FIGURE 1

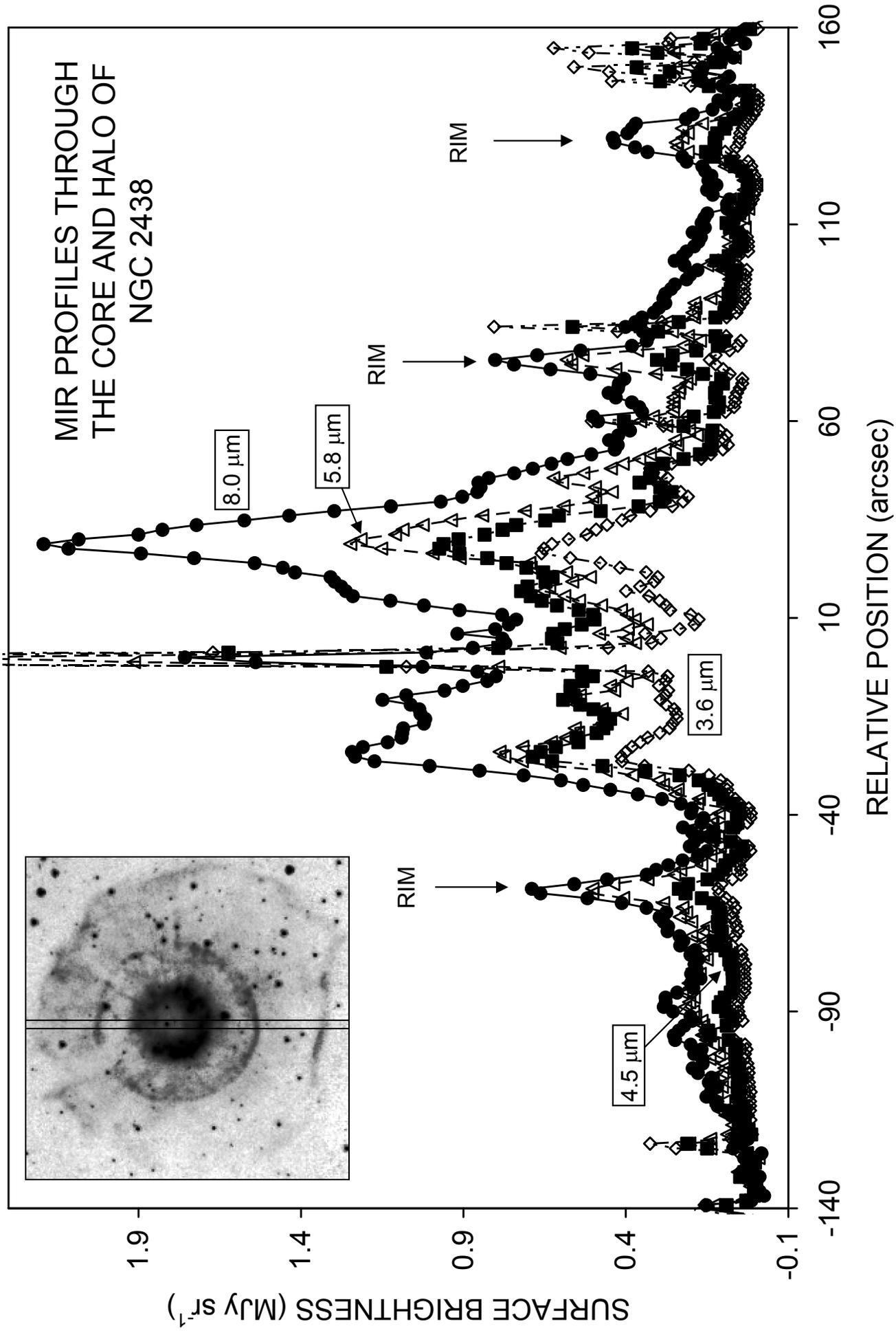

FIGURE 2

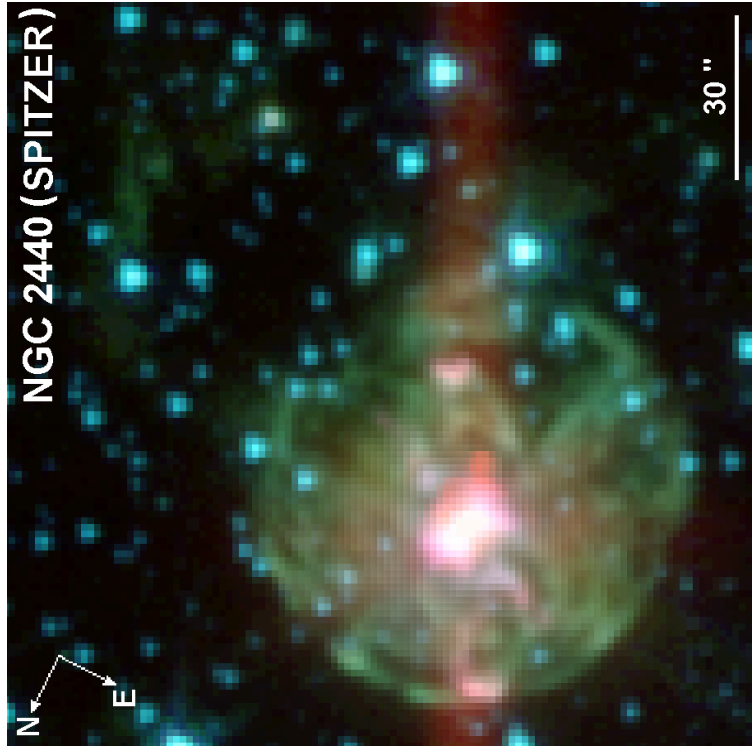
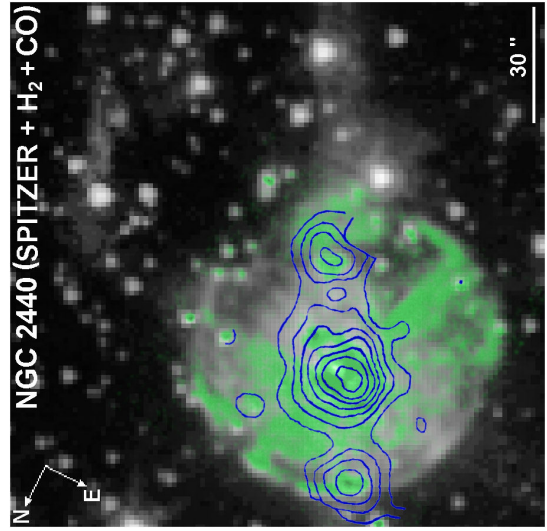
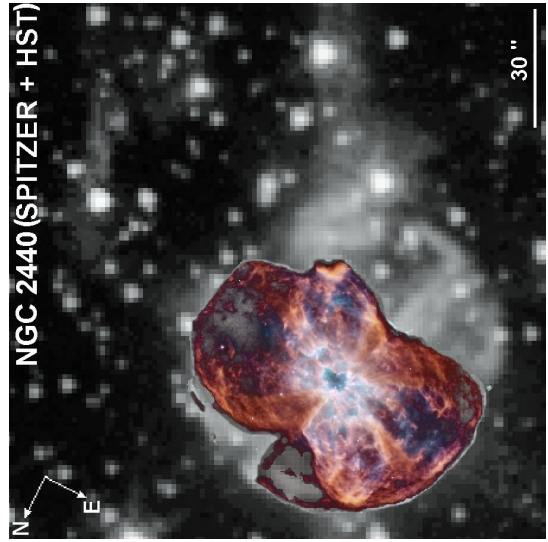

FIGURE 3

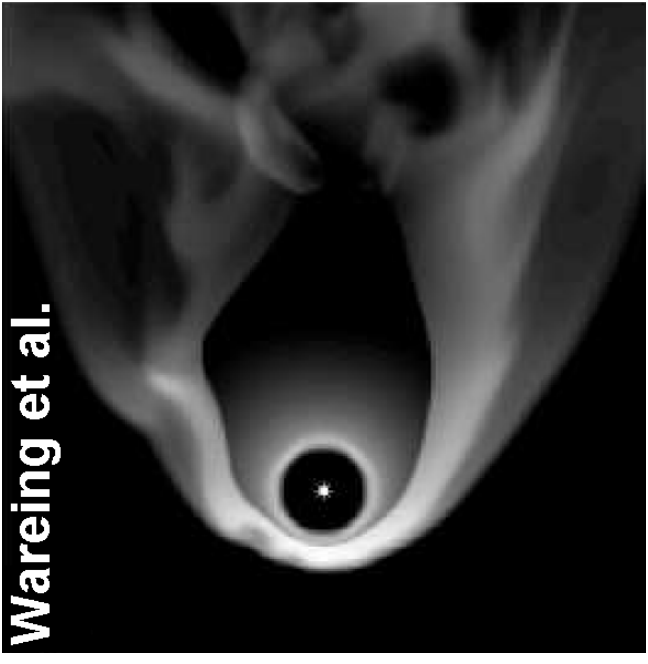
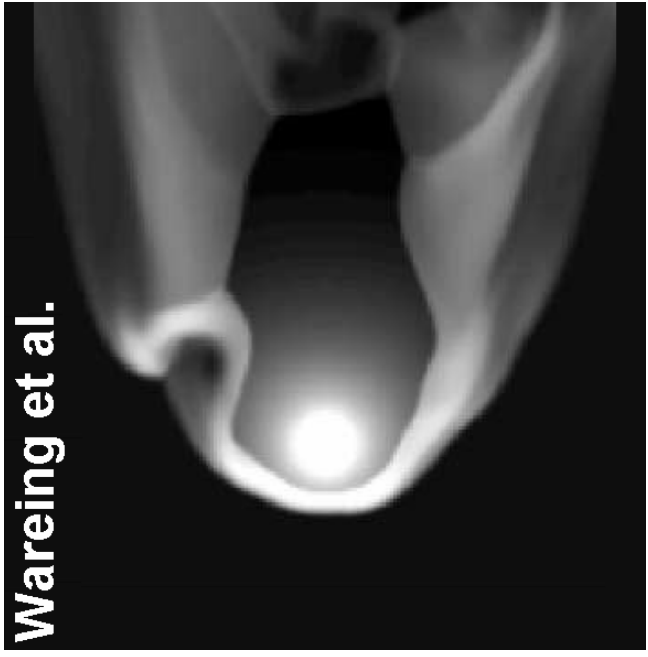
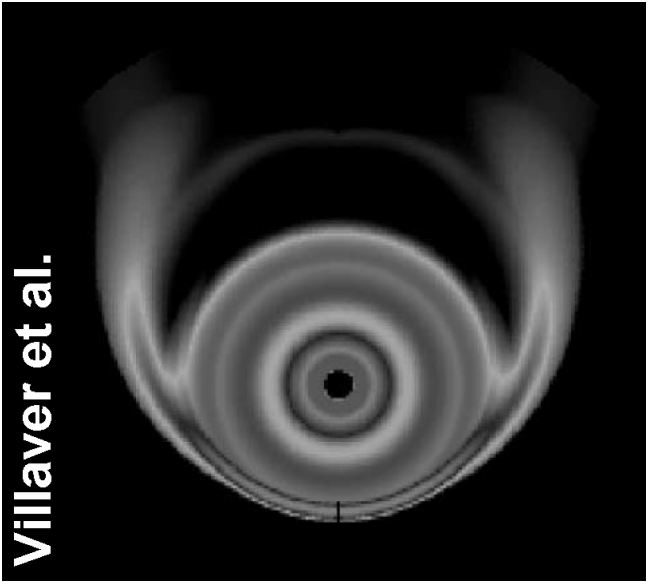
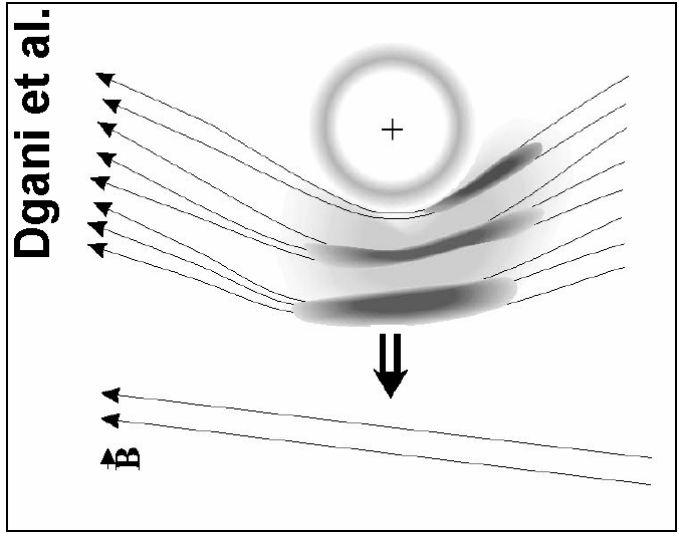
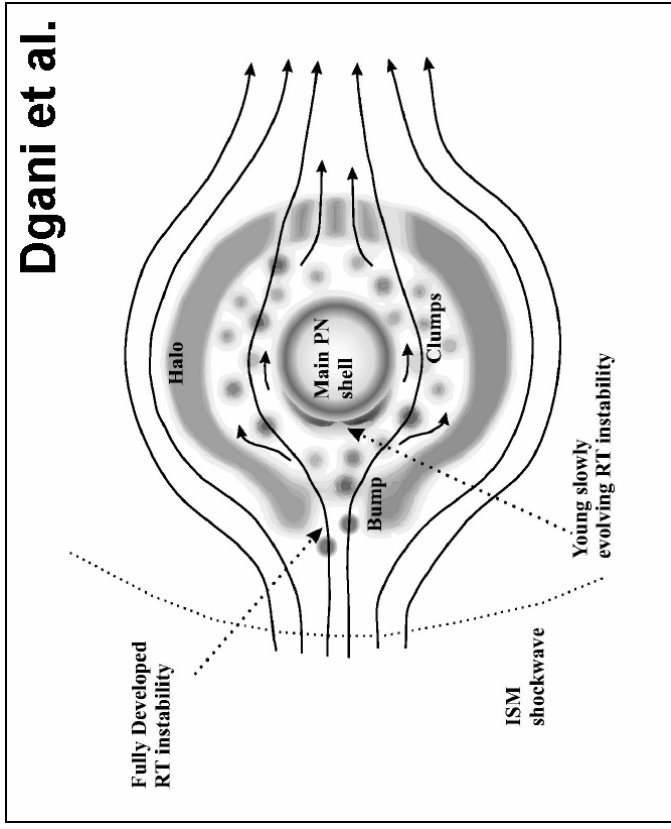

FIGURE 4

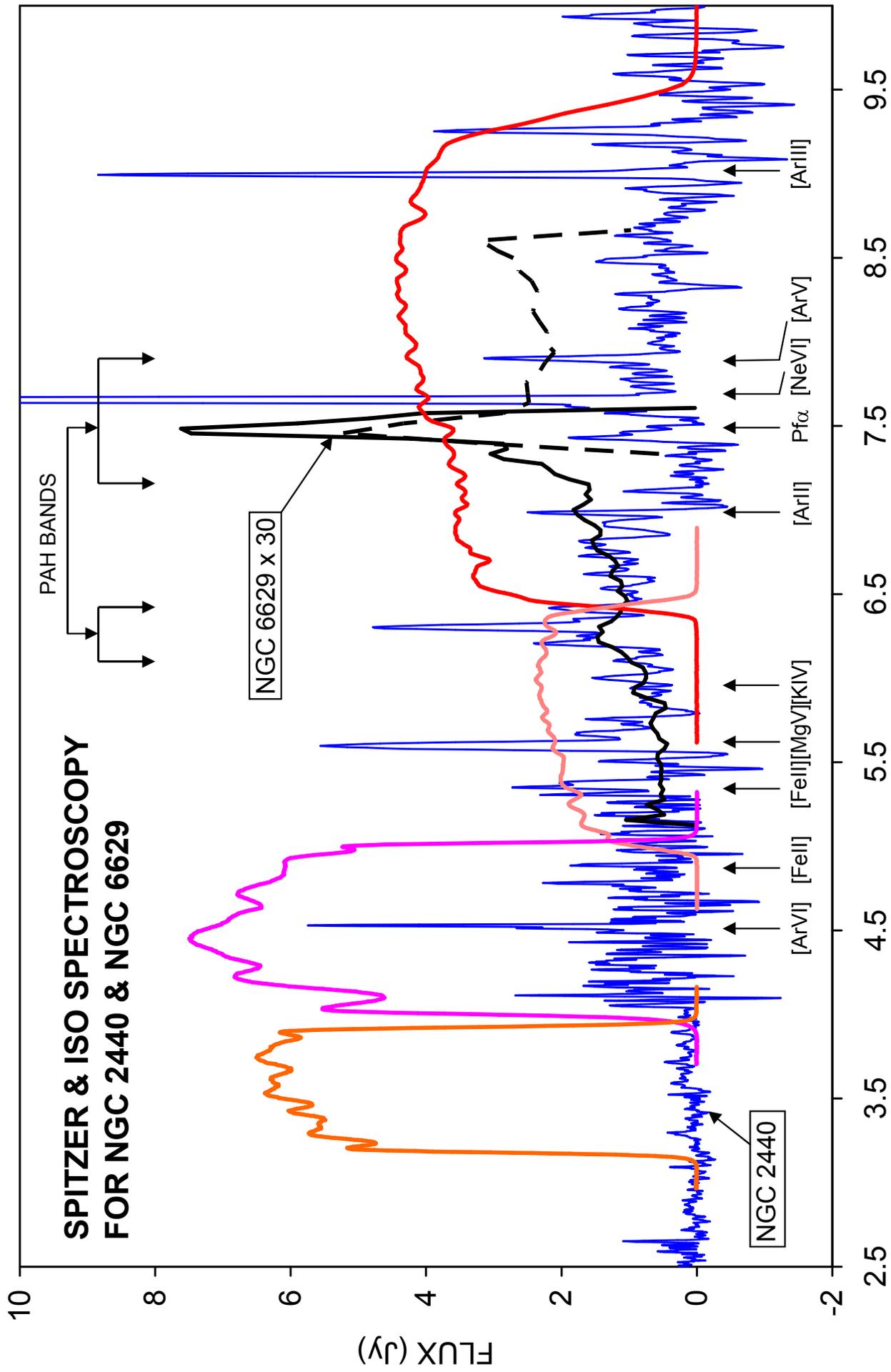

FIGURE 5

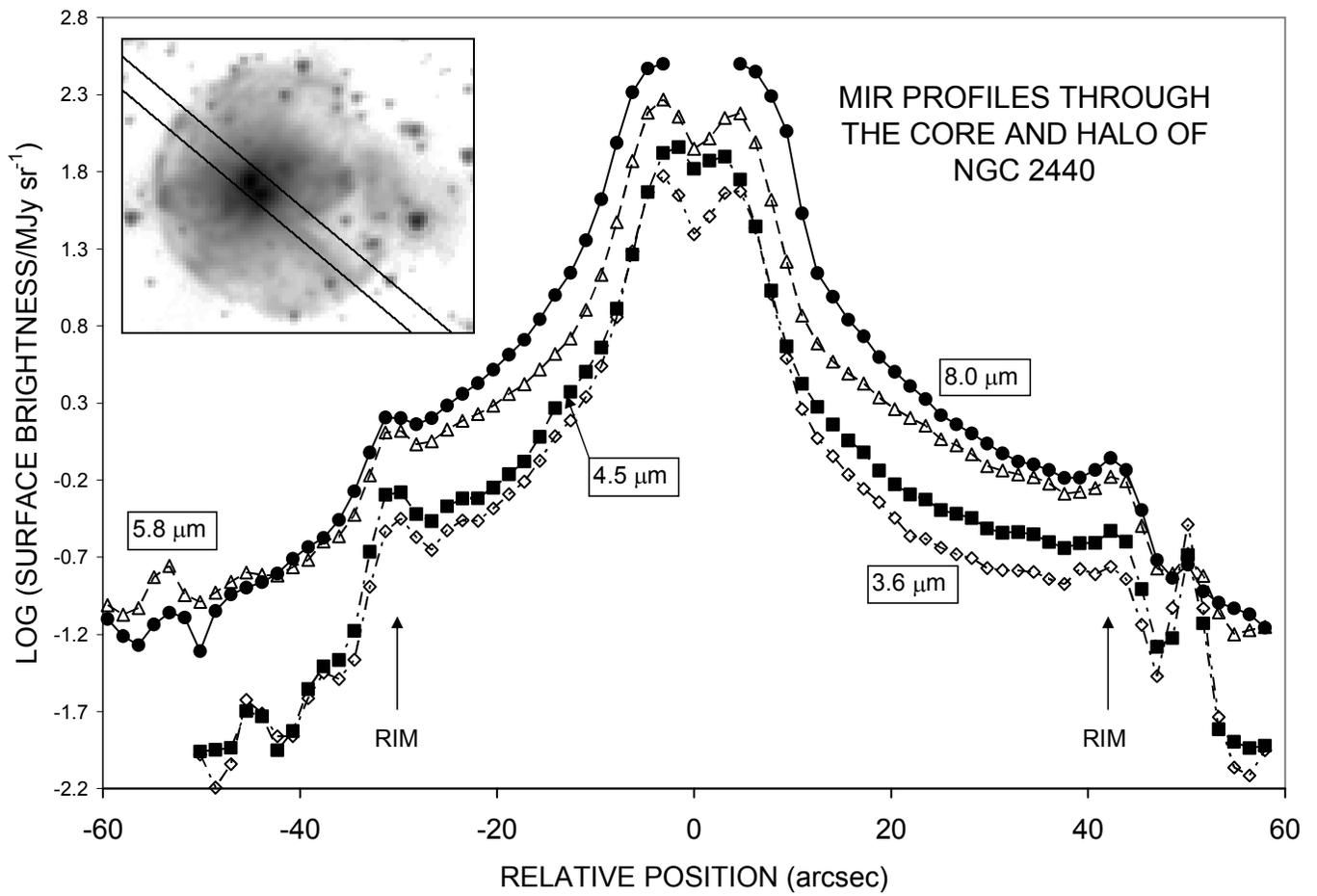
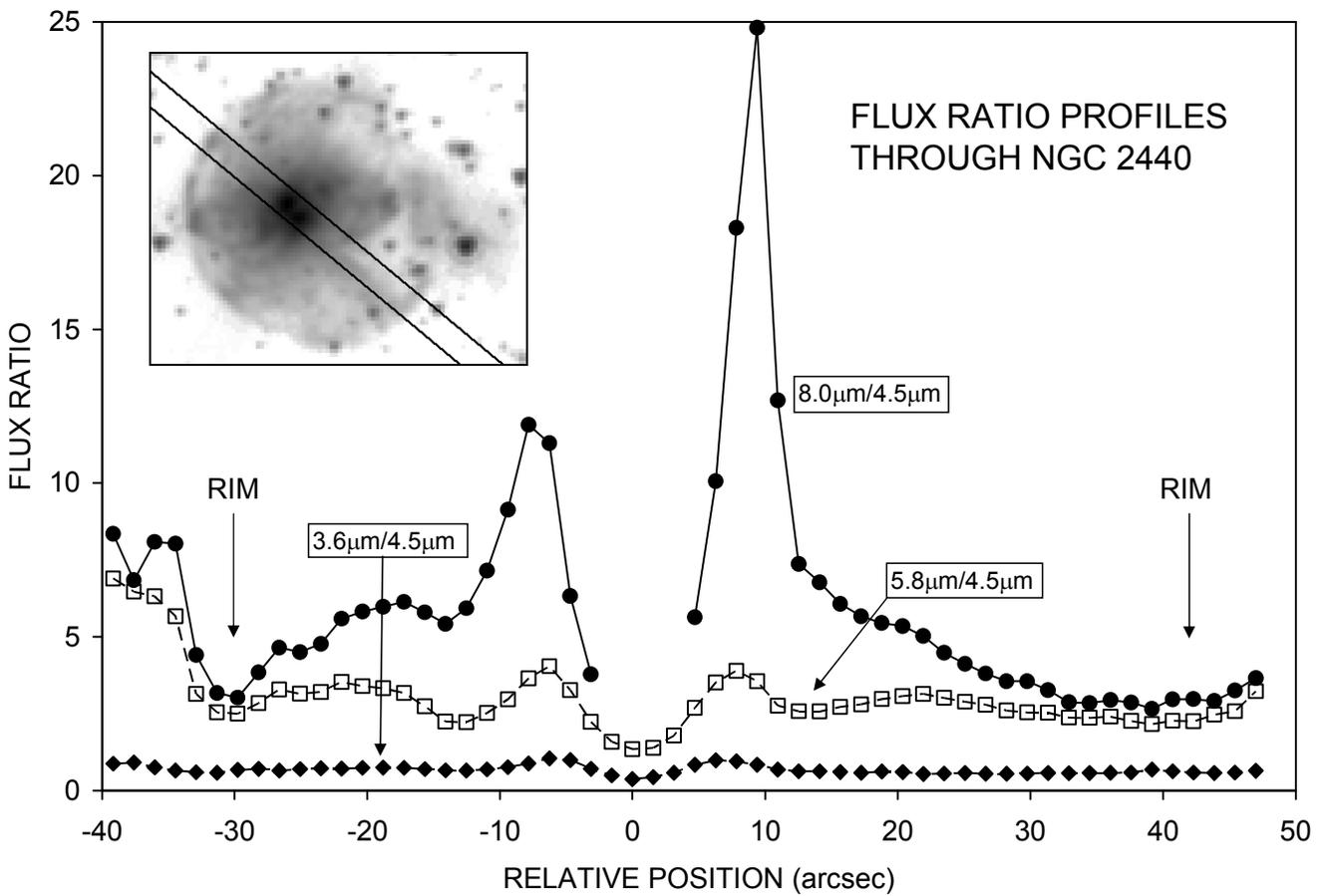

FIGURE 6

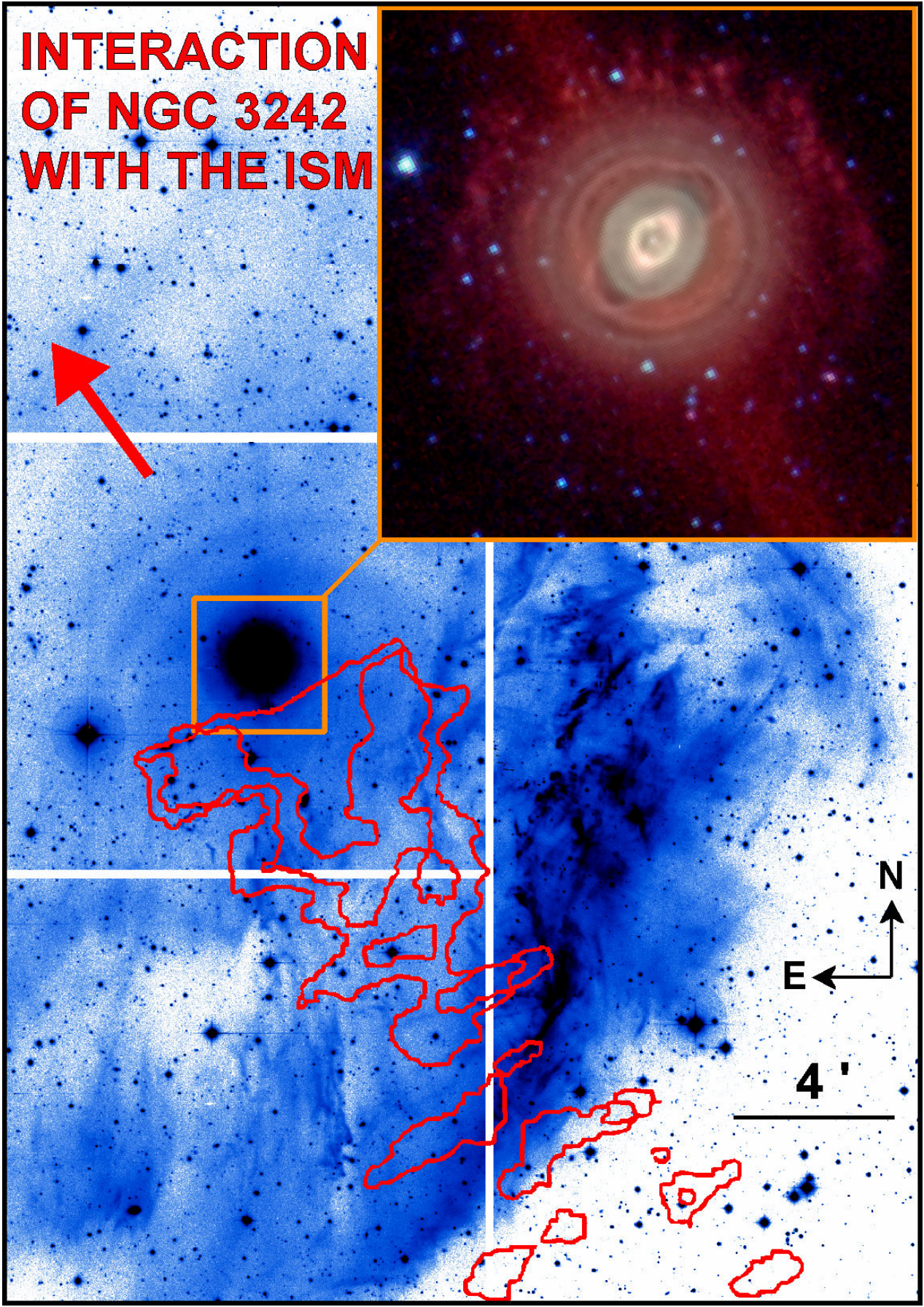

FIGURE 7

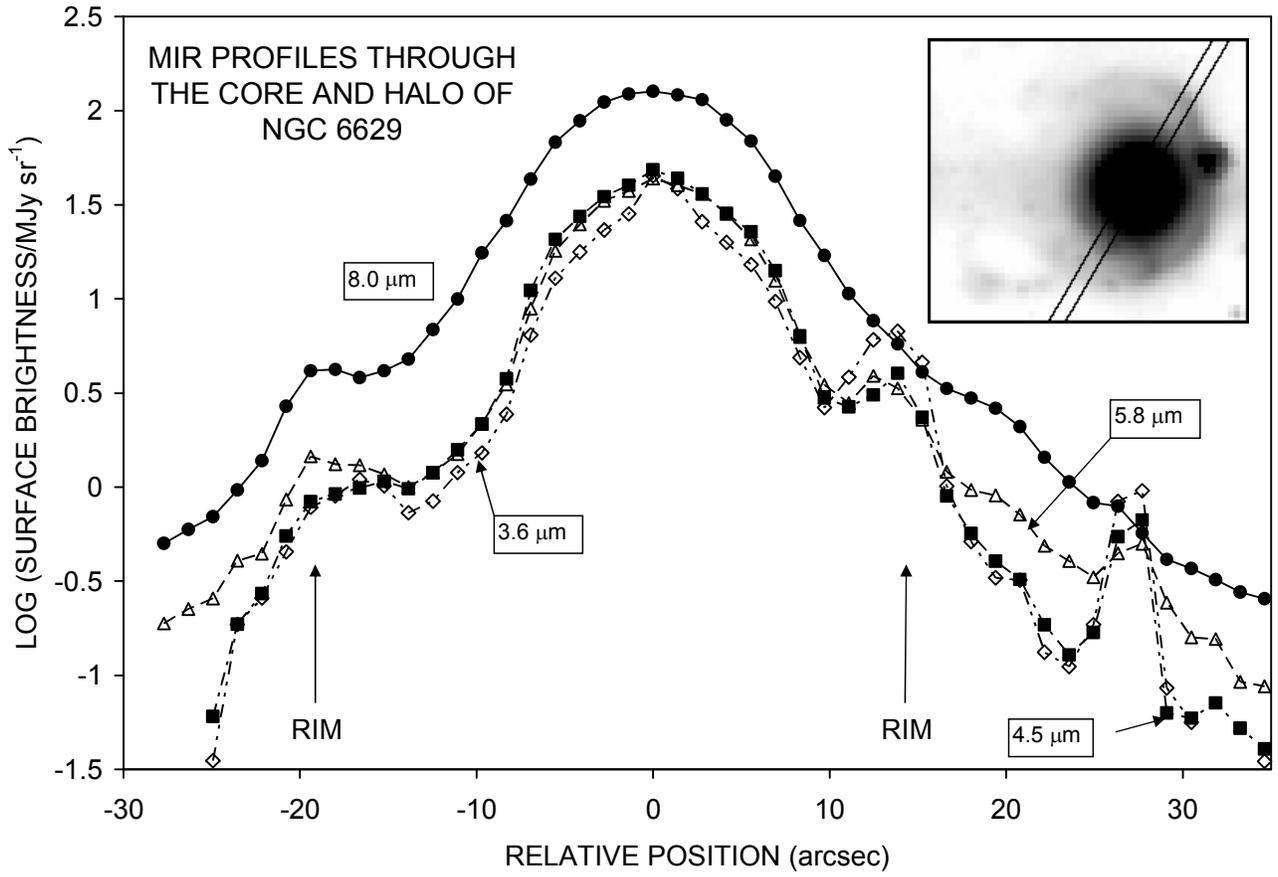
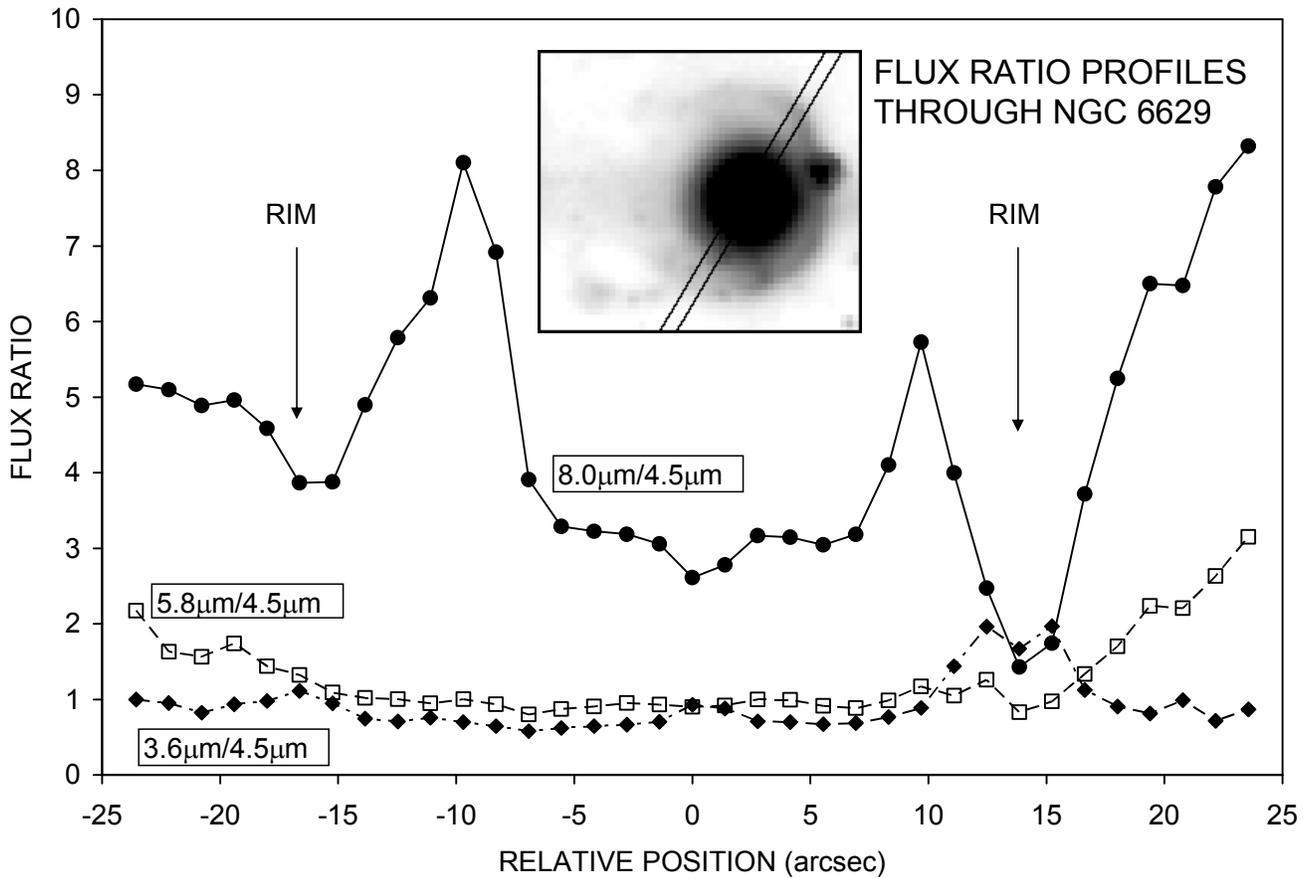

FIGURE 8

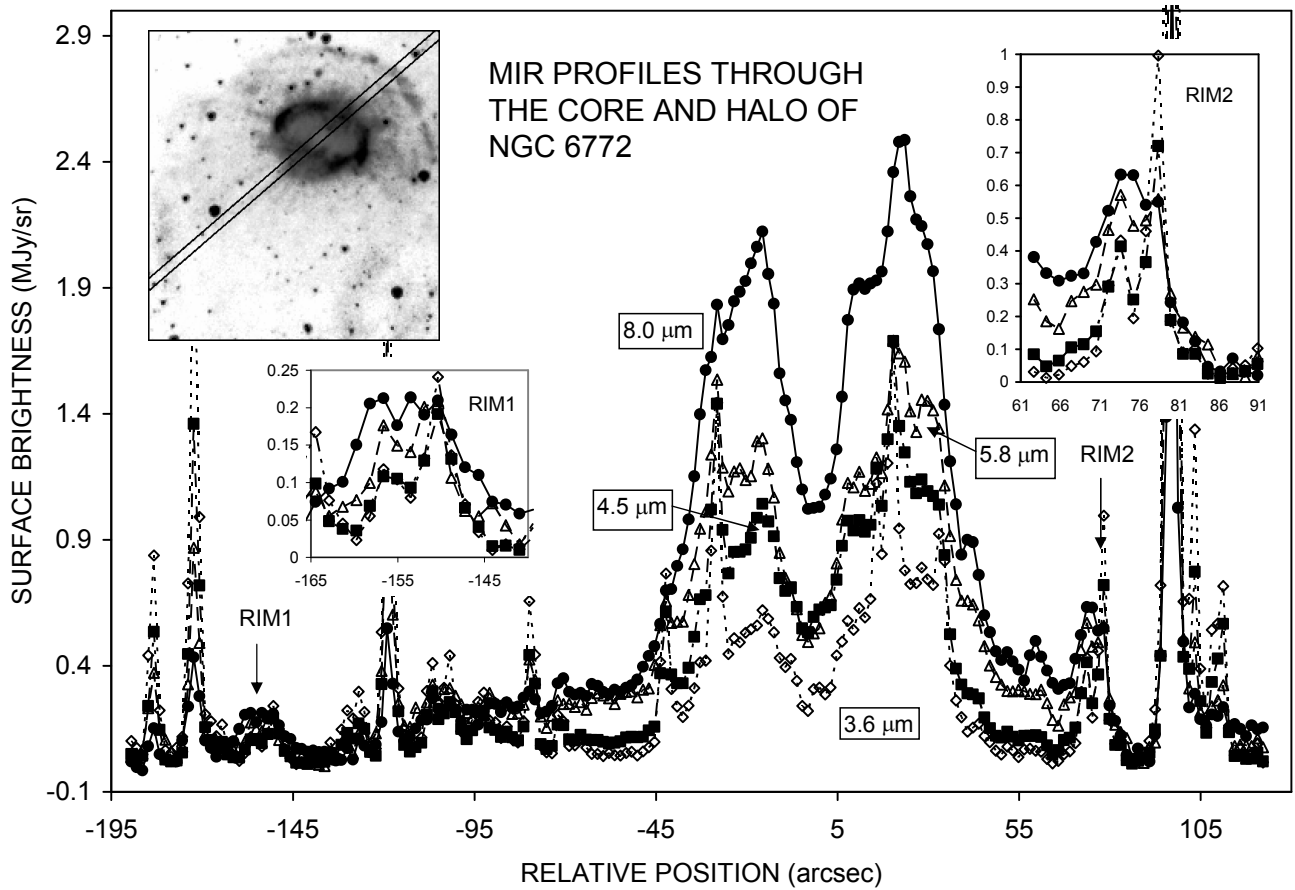
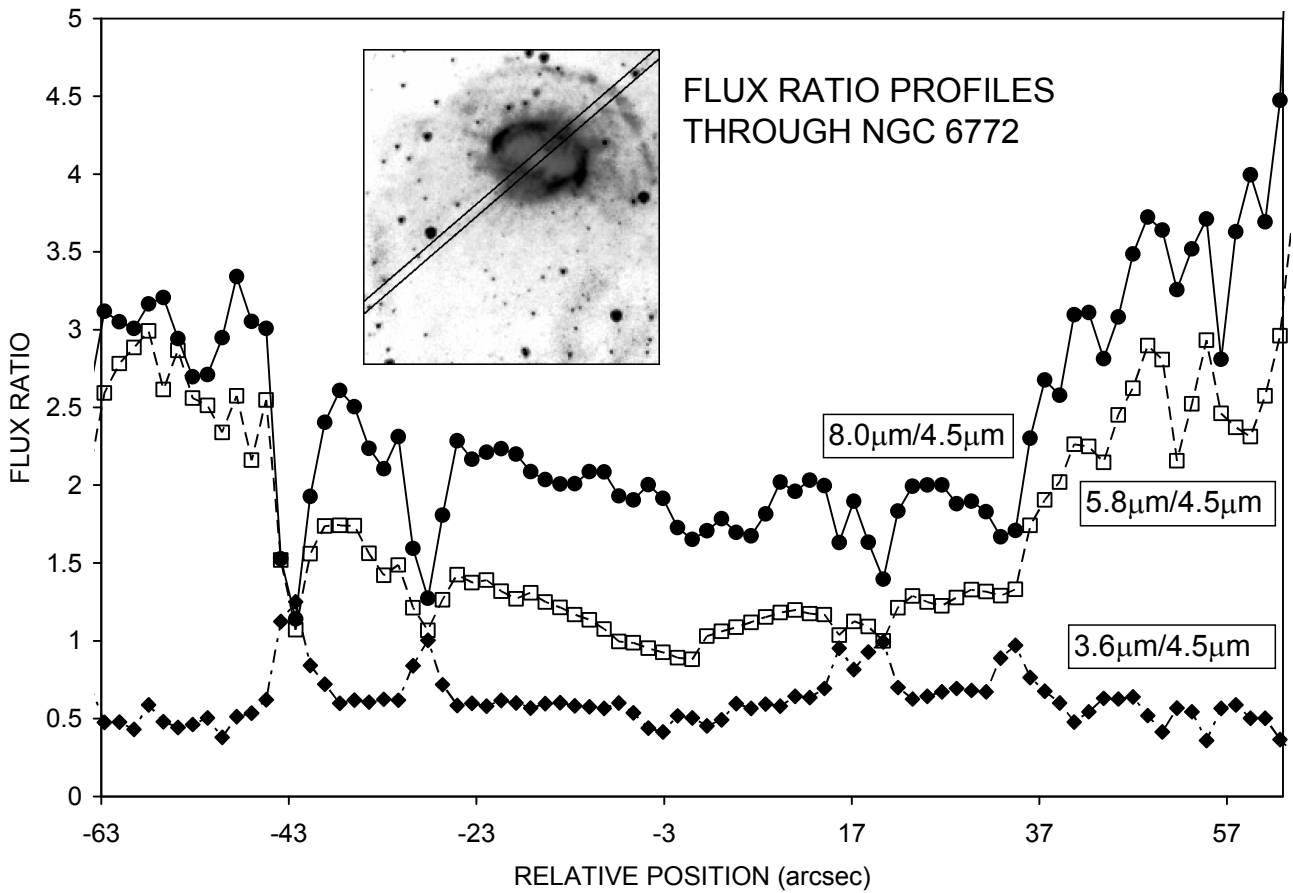

FIGURE 9
42